\documentclass[mathleft
]{an}
\usepackage{graphicx}
\usepackage{times}
\begin{document}

\Pagespan{904}{908}
\Yearpublication{2007}%
\Yearsubmission{2007}%
\Month{10}%
\Volume{328}%
\Issue{9}%
 \DOI{10.1002/asna.200710756}%

\title{Optical flares from the faint mid-dM star 2MASS~J00453912+4140395}

\author{Zs. K\H{o}v\'ari \inst{1}\fnmsep\thanks{Corresponding author:
  \email{kovari@konkoly.hu}\newline}
\and F. Vilardell\inst{2}
\and I. Ribas\inst{3,4}
\and K. Vida\inst{1,5}
\and L. van~Driel-Gesztelyi\inst{1,6,7}
\and C. Jordi\inst{2,4}
\and K. Ol\'ah\inst{1}
}
\titlerunning{Optical flares from the faint mid-dM star 2MASS J00453911+4140396}
\authorrunning{Zs. K\H{o}v\'ari et al.}
\institute{
Konkoly Observatory, P.O.Box 67, H-1525 Budapest, Hungary
\and 
Departament d'Astronomia i Meteorologia, Universitat de Barcelona, c/ Mart\'{\i} i Franqu\`es, 1-11, 08028 Barcelona, Spain
\and
Institut de Ci\`encies de l'Espai -- CSIC, Campus UAB, Facultat de Ci\`encies, Torre C5 - parell - 2a planta, 08193 Bellaterra, Spain
\and
Institut d'Estudis Espacials de Catalunya (IEEC), Edif. Nexus, C/Gran Capit\`a, 2-4, 08034 Barcelona, Spain
\and
E\"otv\"os University, Department of Astronomy, P.O.Box. 32, H-1518 Budapest, Hungary
\and
Observatoire de Paris, section Meudon, LESIA (CNRS), 92195 Meudon Principal Cedex, France
\and
Mullard Space Science Laboratory, University College London, Holmbury St.Mary, Dorking, Surrey, RH5 6NT, UK
}

\received{2007 Feb 19}
\accepted{2007 Mar 5}
\publonline{2007 Oct 18}


\abstract{We present $B$ and $V$ light curves of a large stellar flare 
obtained with the Wide Field Camera at the Isaac Newton 2.5-m telescope 
(La Palma). The source object is a faint ($m_V=21.38$) foreground star in 
the field of the Andromeda galaxy, with its most probable spectral type 
being dM4. We provide an estimate of the total flare energy in the optical 
range and find it to be of the order of $10^{35}$ erg. The cooling phase 
of the large flare shows three additional weak flare-like events, which we 
interpret as results of a triggering mechanism also observed on the Sun 
during large coronal mass ejections.}


\maketitle

\section{Introduction}\label{intro}

Flares are known as sudden and violent events releasing magnetic energy 
and hot plasma from the stellar atmospheres. We observe them on 
magnetically active stars and, much more closely, on the Sun. 
Electromagnetic radiation is emitted across the entire spectrum, from 
radio waves through the optical range to X-rays and $\gamma$ rays. The 
total energy released during a typical solar flare is of the order of 
$10^{30}$~erg, while the largest solar two-ribbon flares can emit up to 
$10^{32}$~erg.

According to the accepted model of solar flares, in the upper atmosphere, 
between oppositely oriented magnetic field lines, a current sheet forms 
and magnetic reconnection takes place, which results in acceleration of 
particles, and thus produces electromagnetic radiation and plasma heating.  
On the basis of the solar paradigm one could simply expect that modelling 
stellar flares is just a question of scaling (e.g., as a function of the 
released energy, size, duration, etc.). However, besides general 
similarities, stellar flare observations also unveil problems that cannot 
be explained with the extended canonical solar flare model, arising from 
the different spectral distribution of the emitted energy, the role of age 
and spectral type of the host star, the multiform magnetic field 
topologies on stars, the tidal forces in active binaries, etc. The only 
way we can come closer to drawing up similarities and differences between 
solar and stellar flares is to continuously broaden the ensemble of 
analyzed flares (solar and stellar as well), thus making an observational 
base of different types of flaring activities available for theoretical 
purposes. This is how solar flare observations can help in the correct 
interpretation of stellar flares, and reversely.

Flare activity in cool stars is a very common phenomenon. Flares in stars 
with spectral types earlier than M are observed mainly in UV and X-rays 
and optical flares (like the most energetic white light flares in the Sun; 
Hudson, Wolfson, \& Metcalf \cite{huetal06}) are rare. However, in the 
less luminous low-mass dM stars optical flares often occur. It is also 
known that red dwarfs from $\approx0.3 M_\odot$ down to the hydrogen 
burning limit can release a considerable amount of energy via flare 
eruptions, even when, in most cases, these objects show very low magnetic 
activity in their quiescent state. In this paper we present photometric 
observations of a large optical flare event observed on 2000 September 25 
that is associated with one such low-mass dM stars. In Sect.~\ref{data} a 
short description of the data is given, and in Sect.~\ref{star} we 
summarize the stellar properties of the host star. In Sect.~\ref{flare} we 
give an estimate of the energy released by the large flare and, finally, 
in Sect~\ref{postflare} an additional two weaker short-term flares are 
identified and a plausible scenario is set up for the three weak 
post-flare events that followed the large flare.

\begin{figure}
 \includegraphics[width=0.48\textwidth]{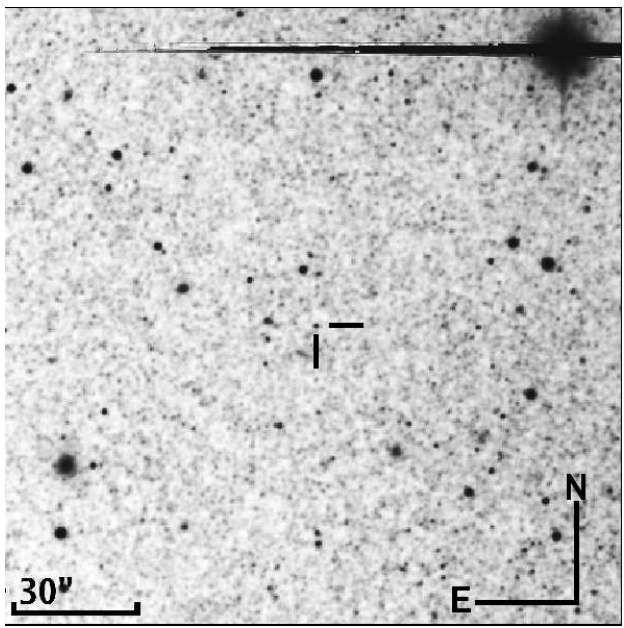}
\caption{Finding chart from the M31 field with our target (INT WFC image).}
\label{finding}
\end{figure}

\begin{table*}
\caption{Catalog data on the observed target}
\label{T1}
\begin{tabular}{lccc}\hline
			& INT WFC & DIRECT &	2MASS \\
\hline
object ID 	& M31\_J00453912+4140395 & D31J004539.1+414039.5 & J00453911+4140396 \\
$\alpha$ (J2000)		& 00\,45\,39.12 &  00\,45\,39.11	& 00\,45\,39.12 \\
$\delta$ (J2000)		& +41\,40\,39.5 & +41\,40\,39.46	& +41\,40\,39.7 \\
$B$			& 23.18 & 	&	\\
$\sigma_B$		& 0.11	&	&	\\
$V$			& 21.38	& 21.62	&	\\
$\sigma_V$		& 0.03	& 0.07	&	\\
$I$			& 	& 18.18	&	\\
$\sigma_I$		& 	& 0.07	&	\\
$J$			& 	&	& 16.35	\\
$\sigma_J$		&	&	& 0.11	\\
$H$			&	&	& 15.98	\\
$\sigma_H$		&	&	& 0.16	\\
$K$			&	&	& 15.63	\\
$\sigma_K$		&	&	& 0.18	\\
\hline
\end{tabular}
\end{table*}

\section{Data}
\label{data}

\begin{figure*}
  \includegraphics[angle=0,width=0.75\textwidth]{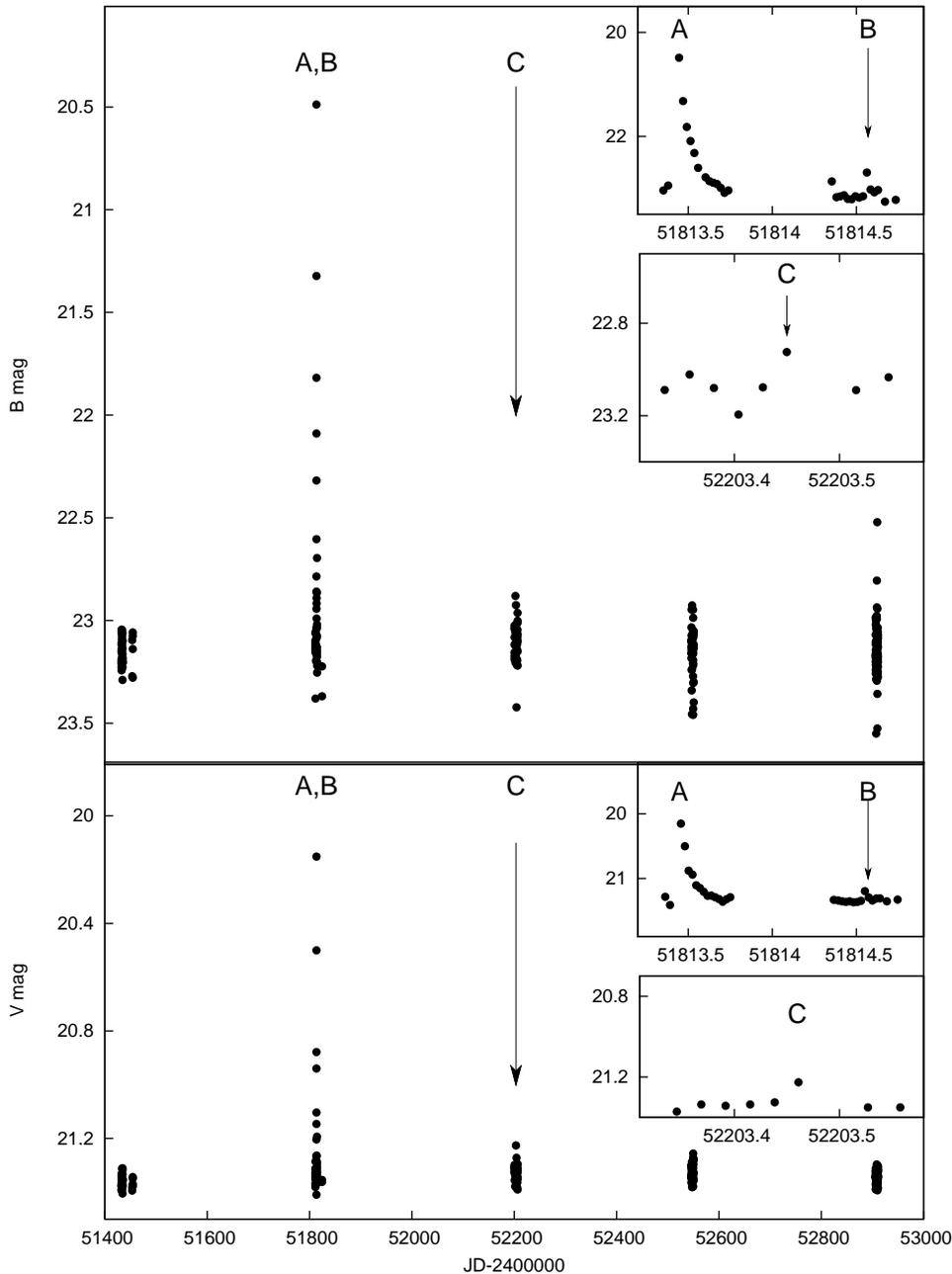}
\caption{Light curves in Johnson $B$ and $V$ bands of 2MASS~J00453912+4140395
taken with the WFC of INT in La Palma, between 1999-2003. Three individual
flare events (flares A, B and C) are marked and zoomed.}
\label{bvobs}
\end{figure*}

Our time-series data were collected between 1999--2003 within the course 
of a variability survey in in the North-Eastern quadrant of the Andromeda 
Galaxy (M31). The observations were acquired with the Wide Field Camera at 
the Isaac Newton 2.5-m telescope (INT) in La Palma (Spain). The program is 
described in detail in Vilardell, Ribas, \& Jordi (\cite{vila06}). In 
brief, the observations, taken during the course of 21 nights, were 
reduced and then analyzed by means of the so-called Difference Image 
Analysis (DIA) technique, which is especially tailored to detect variable 
objects. The resulting data are high precision ($\sim0.01$ mag) 
photometric light curves in the Johnson $B$ and $V$ passbands.  The object 
studied here, which has identifier M31\_J00453912+4140395 in the Vilardell 
et al. (\cite{vila06}) catalog, was flagged during the course of the 
analysis as an object with a large brightening following the 
characteristic light curve shape of a stellar flare. The finding chart of 
the object, also cross-identified as 2MASS~J00453912+4140395, is shown in 
Fig.~\ref{finding}, which contains only a small portion of the full 
32\arcmin$\times$32\arcmin field of view of the WFC. The $B$ and $V$ 
observations resulting from the DIA analysis and subsequent calibration to 
the standard system are plotted in Fig.~\ref{bvobs}, where the large flare 
event (A) is zoomed alongside with two other weaker flares (B, C).

\section{Stellar properties}
\label{star}

Table~\ref{T1} summarizes the quiescent $B$ and $V$ magnitudes from our 
INT data, as well as $V$ and $I$ magnitudes from the DIRECT catalogue
(Mochejska et al. \cite{moch01}), and $J$, $H$ and $K$ magnitudes from the 2MASS 
catalogue (Skrutskie et al. \cite{skru06}). A self-consistent analysis 
using the observed object color indices and the extinction law of Drimmel, 
Cabrera-Lavers, \& L{\'o}pez-Corredoira (\cite{drim03}) indicates a color 
excess of $E(B-V)=0.05$ mag, and, adopting a ratio of 3.1 between $A_V$ 
and $E(B-V)$, leads to an absolute extinction of 0.16 mag. The unreddened 
color indices are most compatible with a spectral type M4 dwarf with 
$M_V=13.8$ mag, $\log L/L_\odot\approx -2.5$, $M\approx 0.16$ M$_\odot$, 
and $R\approx 0.2$ R$_\odot$ according to the tabulations from Baraffe \& 
Chabrier (\cite{bach96}), Baraffe et al. (\cite{bara98}), and Bessell, 
Castelli, \& Plez (\cite{bess98}).  The distance resulting from the 
analysis is of roughly 300~pc.

KPNO plates taken from the M31 field (NOAO Science Archive, Aladin Sky 
Atlas, http:/\slash{}aladin.u-strasbg.fr\slash{}) also agree with a mid-M 
or later spectral type classification since the object is best seen in the 
$I$-band KPNO plates, but also observable in H$\alpha$ and $R$. However, 
in the $U$, $B$, and $V$ band plates our target is indistinguishable from 
the background.

A first step towards evaluating the level of magnetic activity of the 
studied target is to investigate possible cyclic (rotational) variability 
induced by surface inhomogeneities (e.g., cool spots). No convincing sign 
of such modulation was found from our Fourier analysis. Three possible 
explanations can be put forward. First, spot coverage could be too small 
to observe any rotational modulation (this is the case when observing our 
Sun as a star in optical photometric bands). Alternatively, the star could 
be heavily spotted and yet show small or no modulation if spots are evenly 
distributed. If we compare the mean error of the measurements in $B$ is 
0.08 mag with a scatter of 0.11 mag, i.e., there is still room for some 
low-level modulation, although the $V$ band does not seem to show such 
behaviour with a scatter of 0.03 mag and a 0.04 mag formal error. The 
third possible scenario is a low inclination ($i\approx0^{\circ}$) of the 
rotation axis making it not possible to observe any rotational 
variability. However, in this latter case one could expect variability on 
a longer timescale of a few years, because of changes in the overall spot 
coverage according to solar-like spot activity cycle
(see e.g. Ol\'ah, Koll\'ath, \& Strassmeier \cite{olkost00}). No such long-term 
modulation is seen during the 5-year long observing season and 
2MASS~J00453912+4140395 seems to be like the vast majority of cool M 
dwarfs with rotational variability less than $1-2$\% in the visible.

\section{Flare energy estimation}
\label{flare}

Flare statistics of red dwarf stars show that stellar flares can generally 
be divided into two subgroups: a group consisting of relatively small, 
short ($\approx 10^3$ s) impulsive flares, and a more energetic group 
releasing at least $10^{32}$\,erg and lasting $\approx10^4$ s. The latter 
are often related to solar two-ribbon flares which are associated to 
filament eruptions and coronal mass ejections (CMEs). The basic properties 
of flare A observed on 2000 September 25 (beginning at HJD 2451813.45), 
such as the long duration reaching $\Delta t_B\approx 3\times 10^4$ s and 
the large amplitudes of $\Delta B=$2.69 mag and $\Delta V=$1.23 mag 
clearly class this event among the more energetic group.

Flare light curves usually consist of a rapid rise followed by a slower, 
monotonic decay. However, our poorly covered rising phase permits only a 
rough estimate of the physical properties. For the estimation of the flare 
energy first we derived intensity from the magnitude values: 
\begin{equation} \frac{I_{0+f}}{I_0}=10^{\frac{\Delta m_{B,V}}{2.5}}, 
\end{equation} where $I_{0+f}$ and $I_0$ are the intensity values of the 
flaring and the quiescent stellar surfaces, respectively, in one of the 
observed bands. The relative flare energy is then obtained by integrating 
the flare intensity over the flare duration: \begin{equation} 
\mathcal{E}_f=\int_{t_1}^{t_2} (\frac{I_{0+f}(t)}{I_0} - 1) dt. 
\end{equation}

The quiescent stellar fluxes in different bands are estimated assuming a
simple black body energy distribution for a dM4 star with
$T_{\rm eff}=3100$ K and $R\approx0.2$ R$_{\odot}$ (see
Sect.~\ref{star}) from the equation
\begin{equation}
F_\star=\int_{\lambda_1}^{\lambda_2} 4\pi R^2 \mathcal{F}(\lambda) \, S_{B,V}(\lambda) d\lambda,
\end{equation}
where $\mathcal{F}(\lambda)$ is the power function and $S_{B,V}(\lambda)$ is
the transmission function for a given passband. Finally, the total integrated
flare energy is calculated by multiplying the relative flare energy by the
quiescent stellar flux:
\begin{equation}
E_f=\mathcal{E}_f F_\star.
\end{equation}

Since our flare light curve is poorly covered, the abrupt rising phase 
with its real peak value may be estimated as even 1.5--2~mag brighter in 
$B$ and $\approx$1.3~mag brighter in $V$ than the measured maxima. This
assumption corresponds to a photometric flare temperature
of $\approx1.5\times10^{4}$\,K at peak (cf., e.g., Ishida 
\cite{ishi90}; Ishida et al. \cite{ishi91}, and references therein).
According to this, the observed and 
dereddened $B-V$ peak of 0.29~mag is just an upper limit of a more 
reliable $B-V\approx-0.16$~mag.
Calculated luminosities for the quiescent star, and the total flare energy 
for both the minimum fit and the more realistic assumption are summarized 
in Table~\ref{T2}, together with the ratio of the flare energy and the 
quiescent stellar luminosity, often called equivalent flare duration, 
i.e., the time interval in which the star would radiate as much energy as 
the flare itself. The ratio between the flare energies $E_B/E_V=2.01$ (or 
1.83 for the more realistic estimation) is of the order of the statistical 
value of $1.60^{+0.13}_{-0.32}$ from Lacy, Moffett, \& Evans 
(\cite{lacy76}). Using the empirical correlation \begin{equation} 
E_U=1.2\pm0.08E_B \end{equation} we obtain an estimation of 4.27 
(7.61)\,$10^{34}$ erg for $E_U$.  From this we can estimate the total flare 
energy released in the optical range as being of the order of a few times 
$10^{35}$ erg, and resulting in a value of $\simeq10^{36}$ erg for the 
bolometric flare energy (cf. the review of Pettersen \cite{pett89}).
 This 
rough but still realistic estimation (c.f. the example in Pagano et al. \cite{paga97}
with $\sim3$ times more brightness in $U$) indicates that the 2000 September 25 
flare is among the most energetic stellar flares ever observed.

\begin{table}
\caption{Quiescent stellar flux and flare energy estimations. A minimum 
flare energy is estimated both from a minimum fit of the observed data and also
by a (more realistic) reconstruction of the rising phase (values in
parenthesis).}
\label{T2}
\begin{tabular}{lccc}\hline
band	& quiescent flux & flare energy & equivalent duration \\
	& $10^{29}$ [erg/s] & $10^{34}$ [erg] & [$h$]\\
\hline
$B$	&5.09	&	3.56 (6.34) & 	19.4 (34.6)\\
$V$	&12.76	&	1.77 (3.47) & 	3.9 (7.6)\\
\hline
\end{tabular}
\end{table}

\begin{figure*}
  \includegraphics[angle=0,width=0.85\textwidth]{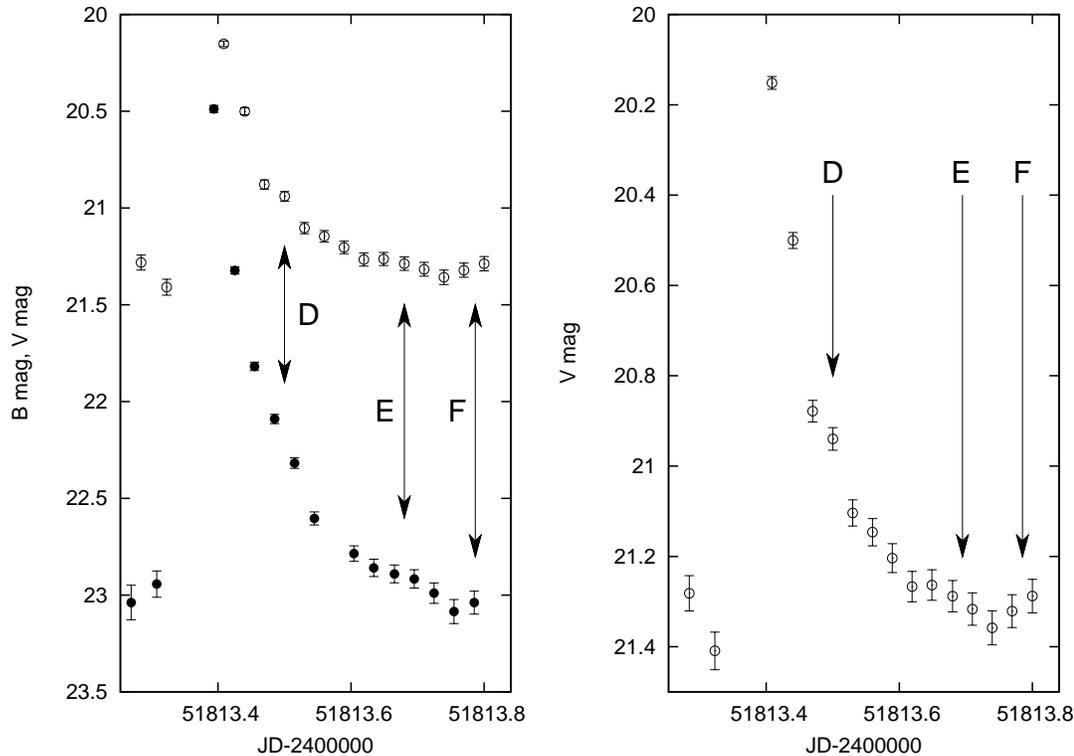}
\caption{The large flare of 2000 September 25 (flare A in Fig.~\ref{bvobs})
with three post-flare events (D,E and F). Left panel shows the $B$ (dots) and
$V$ (circles) light curves together. In the right panel the zoomed $V$ 
curve is plotted alone.} \label{postflarefig} 
\end{figure*}

\section{Weak flaring and post-flare events}
\label{postflare}

There are statistical evidences that flare-like transient phenomena, often 
called microflares, are almost continuously present on M-type dwarfs. 
These events are thought to be one of the major sources of chromospheric 
and coronal heating. We see them (sometimes as nanoflares) also on the Sun 
(e.g., Lin et al. \cite{line84}), however, because of being relatively 
weak and short-term events most of them remain unresolved.

For filtering out such short-duration events from the background of our 
photometric data we searched for small amplitude peaks occurred 
simultaneously in both photometric colours.  We applied different 
filtering algorithms but finally a visual inspection proved to be most 
reliable and efficient. Two such events are apparent. One is just one day 
after flare A at HJD 2451814.55 (marked with B in Fig.~\ref{bvobs}) and 
another one at HJD 2452203.45 (flare C in Fig.~\ref{bvobs}). There are 
some additional albeit less convincing ones, which are only 2-$\sigma$ 
away from the photometric background, and therefore were disregarded. 
Following the method described in Sect.~\ref{flare}, we estimate the total 
energy of flares B and C to be from a few times $10^{32}$ erg up to 
$10^{33}$ erg in $B$, which are characteristic of short-term impulsive 
flares rather than microflares.

When analyzing the decay phase of the light curve of flare A, three other 
short term increases can be identified. Since those events occurred 
simultaneously in both colours, we assume them to be an additional three 
weak flares: one at HJD 2451813.54 (D in Fig.\ref{postflarefig}), another 
one at HJD 2451813.67 (E) and a third one just at the end of the large 
flare light curve at HJD 2451813.74 (F), all of them lasting about $\Delta 
t\approx 8\times 10^3$ s (in the case of flare F, the simultaneous rise in 
$B$ and $V$ together is considered as just the beginning of the flare 
event and its duration is extrapolated). Again, we estimate the total 
flare energies as $2-6\times 10^{31}$ erg in $B$ and about a half of this 
value in $V$. The occurrence of such weak post-flare events in the 
descending phase of a large long duration flare is a reminder of the 
mechanism described first by Attrill et al. \cite{attr07}, which probably 
occurs on the Sun during large CMEs. After energetic solar flares are 
associated with CMEs, the moving footpoints of the blowing up magnetic 
loop crosstalk with opposite polarity flux ropes from the randomly 
distributed magnetic carpet, or other favourably oriented magnetic 
concentrations (active regions), thus forming current sheets and 
triggering other (micro)flare events from time to time. This seems a 
plausible interpretation for the three weak flare-like events (D, E and F) 
observed during the decay phase of flare A.

\acknowledgements

ZsK is a grantee of the Bolyai J\'anos Scholarship of the Hungarian 
Academy of Sciences. ZsK, KV, LvDG and KO are grateful to the Hungarian Science 
Research Program (OTKA) for support under grants T-043504 and T-048961.  
IR, CJ, and FV acknowledge support from the Spanish Ministerio de 
Educaci\'on y Ciencia via grants AYA2006-15623-C02-01 and 
AYA2006-15623-C02-02. This publication makes use of data products from the 
Two Micron All Sky Survey, which is a joint project of the University of 
Massachusetts and the Infrared Processing and Analysis Center/California 
Institute of Technology, funded by the National Aeronautics and Space 
Administration and the National Science Foundation.

\newpage


\end{document}